\documentclass[10pt,conference]{IEEEtran}
\IEEEoverridecommandlockouts
\makeatletter
\def\ps@IEEEtitlepagestyle{%
  \def\@oddhead{\footnotesize Accepted as a conference paper at SANER 2025\hfill}%
  \def\@evenhead{\footnotesize Accepted as a conference paper at SANER 2025\hfill}%
  \def\@oddfoot{}%
  \def\@evenfoot{}}
\def\ps@headings{%
  \def\@oddhead{\footnotesize Accepted as a conference paper at SANER 2025\hfill}%
  \def\@evenhead{\footnotesize Accepted as a conference paper at SANER 2025\hfill}%
  \def\@oddfoot{}%
  \def\@evenfoot{}}
\makeatother
\usepackage{array}
\newcolumntype{R}[1]{>{\raggedleft\arraybackslash}p{#1}}
\usepackage{cite}
\usepackage{float}
\usepackage{amsmath,amssymb,amsfonts}
\usepackage{algorithm}
\usepackage{algpseudocode}
\usepackage{graphicx}
\usepackage{textcomp}
\usepackage{xcolor}
\usepackage{stfloats}
\usepackage{booktabs}
\def\BibTeX{{\rm B\kern-.05em{\sc i\kern-.025em b}\kern-.08em
    T\kern-.1667em\lower.7ex\hbox{E}\kern-.125emX}}
\begin{document}

\title{Cross-System Software Log-based Anomaly Detection Using Meta-Learning\\
% \thanks{Identify applicable funding agency here. If none, delete this.}% 
}
%\author{\IEEEauthorblockN{Anonymous Authors}}

\author{
\IEEEauthorblockN{
1\textsuperscript{st} Yuqing Wang$^1$\IEEEauthorrefmark{2}, 
2\textsuperscript{nd} Mika V. Mäntylä$^1$, 
3\textsuperscript{rd} Jesse Nyyssölä$^1$, 
4\textsuperscript{th} Ke Ping$^1$, 
5\textsuperscript{th} Liqiang Wang$^2$
}
\IEEEauthorblockA{
$^1$Department of Computer Science, University of Helsinki, Helsinki, Finland\\
\{yuqing.wang, mika.mantyla, jesse.nyyssola, ke.ping\}@helsinki.fi
}
\IEEEauthorblockA{
$^2$Department of Computer Science, University of Central Florida, Orlando, United States\\
Liqiang.Wang@ucf.edu
}
\thanks{\IEEEauthorrefmark{2}Corresponding author: Yuqing Wang (yuqing.wang@helsinki.fi).}
}

\maketitle

\begin{abstract}
Modern software systems produce vast amounts of logs, serving as an essential resource for anomaly detection. Artificial Intelligence for IT Operations (AIOps) tools have been developed to automate the process of log-based anomaly detection for software systems. 
Three practical challenges are widely recognized in this field: data labeling costs, evolving logs in dynamic systems, and adaptability across different systems. In this paper, we propose CroSysLog, an AIOps tool for log-event level anomaly detection, considering these challenges. Following prior approaches, CroSysLog uses a neural representation approach to gain a nuanced understanding of logs and generate representations for individual log events accordingly. CroSysLog can be trained on source systems with sufficient labeled logs from open datasets to achieve robustness, and then efficiently adapt to target systems with a few labeled log events for effective anomaly detection. We evaluate CroSysLog using open datasets of four large-scale distributed supercomputing systems: BGL, Thunderbird, Liberty, and Spirit. We used random log splits, maintaining the chronological order of consecutive log events, from these systems to train and evaluate CroSysLog. These splits were widely distributed across a one/two-year span of each system's log collection duration, capturing the evolving nature of the logs in each system. Our results show that, after training CroSysLog on Liberty and BGL as source systems,  CroSysLog can efficiently adapt to target systems Thunderbird and Spirit using a few labeled log events from each target system, effectively performing anomaly detection for these target systems. The results demonstrate that CroSysLog is a practical, scalable, and adaptable tool for log-event level anomaly detection in operational and maintenance contexts of software systems.

%After training, we adapt CroSysLog to Thunderbird and SPIRIT respectively, using a single log split from each system with 2,000 labeled logs, of which 0.2-0.5\% are anomalous for TB and 0.2-0.7\% are anomalous for SPIRIT. Our results show that, after adaptation, CroSysLog achieves FI-scores of 97.55\% for TB and 99.17\% for SPIRIT, when evaluated on each system's 20 distinct log splits, every log split has 10,000 logs of which 0.001-0.5\% are anomalous for TB and 0.001-0.7\% are anomalous for SPIRIT. Compared to baselines, CroSysLog is  

\end{abstract}


\section{Introduction}
\label{sec:introduction}
As modern software systems scale and become more complex, the number of components and services increases and their interactions become more complex, thereby raising the potential for anomalies \cite{zhang2019robust}. Detecting anomalies promptly is essential to prevent them from escalating into significant problems or causing system failures, thus maintaining system reliability and performance \cite{zhang2024metalog}. Logs, which provide records of system operations during runtime, serve as an essential source of information for anomaly detection \cite{he2016experience}. However, the sheer volume of log data and their variations across different parts of the system make manual log analysis both challenging and time-consuming \cite{lin2016log,le2022log}.

Artificial Intelligence for IT Operations (AIOps) tools have been
developed to automate the process of log-based anomaly detection
for software systems by using machine learning methods. However, there are three practical challenges widely recognized in this field: 

%However, our empirical study on existing AIOps tools (Section \ref{sec:related_work}) indicates that no single tool is designed to address three practical challenges together:

\begin{enumerate}
    \renewcommand{\labelenumi}{\textbf{C\theenumi.}}
    \item \textbf{Data labeling cost.}  Labeling log data requires domain expertise and manual effort%. Experts must leverage their understanding of software architecture, operational behaviors, and potential failure modes to analyze vast amounts of log data, labeling anomalies 
    \cite{wittkopp2021loglab,yang2021semi}. The task is further complicated by the complexity of distributed systems, where the root cause of an issue may span multiple interconnected components \cite{fu2009execution,zhou2018fault}.

    \renewcommand{\labelenumi}{\textbf{C\theenumi.}}
    \item \textbf{Log evolution.} Software systems evolve according to laws of software evolution\cite{lehman1979understanding}, and as a result, software logs also change over time \cite{zhang2019robust, kabinna2018examining,li2020swisslog}. For example, a study observed that
    20\%-45\% of log statements changed in the studied projects \cite{kabinna2018examining}; another study \cite{he2016experience} reported that Google systems generate up to thousands of new logs each month. %This poses a challenge to maintaining the performance of trained AIOps tools within a system over time. %The need for efficient continuous adaptation of AIOps tools to quickly evolving logs is necessary to maintain reliable performance.

    \renewcommand{\labelenumi}{\textbf{C\theenumi.}}
    \item \textbf{Cross-system adaptability:} Software systems vary in the logs they generate \cite{he2016experience,cheng2023ai}.  The AIOps tool trained on one system's logs will most likely struggle to perform well on another system's logs \cite{cheng2023ai,chen2020logtransfer}. However, organizations often run numerous software systems, while training a separate AIOps tool for each system can be costly and time-consuming \cite{he2016experience}.
\end{enumerate}

Some supervised AIOps tools are specifically designed to address C2 by adapting to evolving logs; however, they still face C1 and C3 because they need a lot of labeled log data for training and often fail to work well on logs from different systems. Unsupervised AIOps tools address C1 by learning from unlabeled log data. Despite this advantage, they tend to be less effective than supervised tools \cite{xie2024logsd,zhang2019robust} and  they are trained on the characteristics of log data within a system and lack mechanisms for knowledge retention and transfer to other systems (C3). Recently, several studies \cite{zhang2024metalog,chen2020logtransfer,han2021unsupervised, hashemi2024onelog}  have considered C1-C3 together and thus proposed transfer learning-based tools. Such tools can be trained on a source system with sufficient labeled logs and applied to new contexts (using a few or no labeled logs), i.e., adapting to new logs generated by the system it was trained on or logs from entirely different systems. 

Our study extends prior transfer learning studies in response to C1-C3. In this paper, % our study extends the prior  transfer learning solution to further tackle the practical challenges C1-C3 in complex software operational and maintenance contexts. 
we propose CroSysLog, an AIOps tool for log-event (i.e., log line) level anomaly detection (i.e., detect anomalous log events). 
%CroSysLog can be trained on labeled logs from source systems and efficiently adapt to new target systems with a few labeled data to perform effective anomaly detection. 
%The motivation behind developing CroSysLog stems from 
We found that many open datasets containing well-labeled logs from real-world software systems are available, see Section \ref{sec:empirical_study}. 
CroSysLog can be trained on sufficient labeled logs from source systems in these open datasets to achieve robustness, and then quickly adapting to target 
%dynamic, %real-world 
systems with a few labeled logs for effective anomaly detection. This reduces the need for extensive data labeling in target systems (C1). Once trained, CroSysLog can adapt to different target systems to perform anomaly detection effectively (C3), and adapt to evolving logs within these systems using only a few newly labeled logs.

%Once trained, CroSysLog adapt to different target systems and effectively perform anomaly detection~(C3);  CroSysLog can adapt to eveolling logs, with a few labled logs,  

CroSysLog detects anomalies at the log-event level. Many AIOps tools detect anomalies at the log-session (i.e., log sequence) level, analyzing a sequence of logs over each session to determine whether each session as a whole is anomalous or normal \cite{hashemi2024onelog}. The session level method has proven effective for systems where logs are organized by session identifiers, e.g., ``Block IDs'' in HDFS (a Hadoop distributed file system) \cite{zhang2019robust}. 
In some systems, logs are generated by different components without direct session identifiers. Prior studies organized logs from such systems into fixed windows to detect whether the entire window is anomalous or not \cite{zhang2024metalog,xie2024logsd}, however, further root cause analysis (RCA) is still required to pinpoint which log events are abnormal within each anomalous window in order to identify the responsible or affected system components. CroSysLog extends this by directly detecting which log events are anomalous within each fixed window, facilitating more accurate and quicker RCA in such systems.

CroSysLog employs a piror neural representation approach \cite{le2021log} to gain a nuanced understanding of logs and generate log representations for individual log
events accordingly. CroSysLog uses a Long Short-Term Memory network (LSTM) as the base model to analyze each log event within its fixed-window context and detect whether it is anomalous or not. We train CroSysLog using a meta-learning algorithm MAML \cite{finn2017model}, which equips the base model LSTM with robust parameters for cross-system adaptability by learning from varied labeled log events across multiple source systems.

We evaluate CroSysLog using open datasets of four large-scale distributed supercomputing systems: BGL, Thunderbird, Liberty, and Spirit. In these datasets, logs are produced by different system components without direct session identifiers. We train CroSysLog using sufficient labeled log events from Liberty and BGL as source systems. After training, we adapt CroSysLog to Thunderbird and Spirit as target systems separately, using a few labeled log events from each system. %, using each system's random single log split with 2,000 labeled consecutive log events, of which 0.2-0.5\% are anomalous for Thunderbird and 0.2-0.7\% are anomalous for Spirit. 
After adaptation, CroSysLog achieves F1-Scores of 97.55\% for Thunderbird and 99.17\% for Spirit, when evaluated on each system's 20 distinct, random log splits. These splits were widely distributed across a one/two-year span
of each system’s log collection duration, capturing the
evolving nature of the logs in each system.
%Each log split contains 10,000 consecutive log events with 0.01-0.5\% anomalies for Thunderbird and 0.01-0.7\% for Spirit.
CroSysLog achieves similar high F1-scores as the most effective one-to-one transfer learning-based tool (trained from a single source system to a single target system). CroSysLog outperforms supervised tools that excel at analyzing evolving logs, as they only achieve F1-Scores ranging from
24.75\% to 72.30\% on target systems, when trained using the same number of labeled log events as CroSysLog from each target system. Compared to most baselines, CroSysLog requires at least three times less training time and uses less than 60\% of the testing time. We have made  CroSysLog's code publicly available\footnote{https://github.com/yuqwang/CroSysLog}. In summary, our main contributions are:
\begin{itemize}
    \item We propose CroSysLog, an AIOps tool for log-event level anomaly detection in software systems. It outperforms baseline approaches by efficiently learning from varied logs of source systems and adapting to target systems with a few labeled logs for effective anomaly detection.
    \item CroSysLog is the first many-to-many transfer learning tool, capable of efficiently adapting to multiple target systems for effective anomaly detection.  
    \item Evaluation results on open datasets demonstrate the performance of CroSysLog. 

\end{itemize}

\section{Background}
\subsection{Empirical study on software logs in open datasets}
\label{sec:empirical_study}

Many open software log datasets are available for different software systems. For example, Loghub \cite{zhu2023loghub} collects log datasets from 19 software systems, e.g., distributed systems, operating systems, mobile systems, server applications, and standalone software; Usenix CFDR repository \cite{oliner2007supercomputers} provides log datasets from five high-performance computing systems. 

Our study uses log datasets of four large-scale distributed supercomputing systems, BGL, Thunderbird, Liberty, and Spirit, from Usenix CFDR repository to build our dataset. %BGL contains logs collected from a BlueGene/L supercomputer system at Lawrence Livermore National Labs in Livermore, with 131,072 processors and 32,768GB of memory. Thunderbird contains logs from a Thunderbird supercomputer system at Sandia National Labs in Albuquerque, with 9,024 processors and 27,072GB of memory. Liberty contains logs from a Liberty supercomputer at Sandia National Labs in Albuquerque, with 512 processors and 944 GB of memory. Spirit contains logs from a Spirit supercomputer at Sandia National Labs in Albuquerque, with 1028 processors and 1024 GB of memory. 
In these four datasets, each log event is labeled as either normal or anomalous.
Table \ref{tab:stat_datasets} shows log statistics for these datasets. Prior empirical studies \cite{zhang2019robust,li2020swisslog,landauer2024critical} have observed the evolving nature of logs in these open datasets, aligning with our identified characteristic C2. Besides, we identified other characteristics of log data in these datasets:

\begin{table*}[ht]
\centering
\caption{Statistics of our dataset}
\label{tab:stat_datasets}
\begin{tabular}{l  r r r r r r r r r}
\toprule
% Header row
  & & \multicolumn{2}{c}{BGL} & \multicolumn{2}{c}{Thunderbird} & \multicolumn{2}{c}{Liberty} & \multicolumn{2}{c}{Spirit}  \\
 \cmidrule(l){3-10}
% Sub-header row
  & & Abs. &  Rel. & Abs. &  Rel. & Abs. &  Rel. & Abs. &  Rel.\\ 
 \midrule
Number of log events & Total & 4,747,963 & 100.0\% & %211,210,936 
211,212,174 & 100.0\% & 265,569,231 & 100.0\% & %272,298,895 
272,298,969 & 100.0\% \\

 & Normal & 4,399,503 & 92.7\% & %193,743,477
 193,744,715 & 91.7\% & 168,478,453 & 63.4\% & 90,656,272 & 33.3\% \\
 
  & Abnormal & 348,460 & 7.3\% & 17,467,459 & 8.3\% & 97,090,778 & 36.6\% & 181,642,697 & 66.7\% \\
  \hline

% Number of unique  & Total & 302 & 100\% &  &  & 1287 & 100\% &  1337 & 100\% \\
% log lines & Normal & 256 & & & & 1269 & & 1294 &  \\
% & Abnormal & 46 & & & & 18 & & 43 & \\

% \hline

 Unique anomaly labels & & \multicolumn{2}{c}{41} & \multicolumn{2}{c}{34} & \multicolumn{2}{c}{22} & \multicolumn{2}{c}{32}  \\
 \cmidrule(l){1-10}

 Collection duration & & \multicolumn{2}{c}{2005-2006 (214 days)} & \multicolumn{2}{c}{2005-2006 (244 days)} & \multicolumn{2}{c}{2004-2005 (575 days)} & \multicolumn{2}{c}{2005-2006 (556 days)}  \\
 \cmidrule(l){1-10}

\end{tabular}
\vspace{-10pt}
\end{table*}

\textit{A1. Log distribution variance: } %In some systems, normal logs significantly outnumber anomalous ones, whereas in others, the situation is reversed. 
 In software systems, the distribution of logs can vary significantly. % Spirit features a higher proportion of anomalous logs, while in other datasets, the majority of log entries are normal. 
 As shown in Table~\ref{tab:stat_datasets},  BGL and Thunderbird have less than 9\% anomalous log events, Liberty has 36.6\% anomalous log events, Spirit has 66.7\% anomalous log events.  Besides, in these datasets, normal and anomalous logs tend to aggregate in clusters rather than being uniformly distributed. For instance, in BGL,  anomalous logs tend to cluster at specific times, e.g., continuous anomalous logs were generated due to an ``instruction cache parity error" on June 3, 2005, and spanned many entries; in Spirit, many anomalous logs are clustered around several specific system failures, with normal logs typically appearing intermittently between these clusters, indicating periods of recovery or stabilization following critical anomalies.

\textit{A2. Log source difference:} In these datasets, logs are generated by different system components and are often intermixed. Each log event uses the system-specific terminology to describe its source (i.e., the originating system component). We describe several examples in each system here. In BGL, ``MMCS'' indicates log events related to the machine mode control system; ``APP'' refers to log events generated by the applications running on the supercomputer. In Thunderbird, "CHK DSK" indicates logs that refer to events where the system performs disk integrity checks. In Liberty, ``pbs\_mom'' refers to log events related to the portable batch system for job scheduling; ``sshd'' log events are generated by the secure shell daemon. In Spirit, ``dhcpd'' refers to log events related to a DHCP server daemon; ``syslog-ng'' related log events are generated by the local logging daemon.

\textit{A3. Log content diversity and variation:} Table \ref{tab:example_logs} shows example logs randomly chosen from each system in our dataset, showing this characteristic. The content of logs exhibits both diversity and variation, not only across different components within a system but also across different systems. These differences stem from the unique operational characteristics and status reports of each system and its components, leading to a broad spectrum of log content details that enhance the complexity of log analysis and interpretation. 

\begin{table}[ht]
\centering
\caption{Example logs from the supercomputers.}
\label{tab:example_logs} 
% Table 1
\begin{tabular}{p{8.3cm}}
\toprule
\textbf{BGL}  \\
KERNEL INFO	instruction cache parity error corrected \\
KERNEL	FATAL	machine check interrupt  \\
APP	FATAL ciod: Error loading path: invalid or missing program image, No such file or directory \\
MMCS INFO ciodb has been restarted. \\ \midrule

\textbf{Thunderbird} \\
kernel: Losing some ticks... checking if CPU frequency changed \\
pbs mom: Connection refused (111) in open demux, open demux: cannot[...]  \\
Accepted publickey for root from {IP address} port 44278 ssh2 \\
session opened for user root by (uid=0)  \\
check-disks: [node:time] , Fault Status assert [...] \\ \hline

\textbf{Liberty}\\ 
pbs\_mom Bad file descriptor (9) in wait\_request, select failed \\
pbs\_mom, wait\_request failed \\
sshd connection from "\#1335\#" \\
sshd Local disconnected: Connection closed by remote host. \\
\midrule

\textbf{Spirit}\\
kernel: cciss: cmd 0000010000a60000 has CHECK CONDITION\\
DHCPREQUEST for {IP address} from {Mac address} via eth1: unknown lease {IP address}. \\
STATS: dropped 42152 \\
syslog-ng startup succeeded \\
Changing permissions on special file /dev/logsurfer \\ 
\bottomrule
\end{tabular}
\vspace{-10pt}
\end{table}

\subsection{Related work and existing approaches}
\label{sec:related_work}
We explored existing AIOps tools aimed toward addressing practical challenges C1-C3, extend Section \ref{sec:introduction} to further examination of approaches deployed by these tools.
%elated work that has made contributions toward addressing the challenges we identified: C1 (Data Labeling Cost), C2 (System Dynamics and Log Evolution), and C3 (Generalization Across Different Systems). We describe these studies in detail below. 

 Several supervised tools,  including LogRobust \cite{zhang2019robust}, NeuralLog \cite{le2021log}, SwissLog \cite{li2020swisslog}, and a robust and transferable framework (RTF) \cite{ott2021robust}, focus on robust log-session level anomaly detection in dynamic systems with evolving logs, specifically addressing C2. LogRobust and SwissLog utilize an attention-enhanced bidirectional LSTM (Bi-LSTM) to analyze log sequences and weight critical features within log sequences, improving the adaptability to emerging patterns and anomalies as logs evolve. NeuralLog, SwissLog, and RTF use pre-trained models (e.g., BERT, GPT) to extract  semantic information from logs to generate log representations, enabling a more nuanced understanding of evolving logs. NeuralLog omits the traditional log parsing process \cite{he2017drain} and uses BERT with subword tokenization on raw logs (after removing non-characters), improving the handling of new, changing, and Out-Of-Vocabulary (OOV) words in evolving logs \cite{le2021log}. 
 
 %It employs a transformer-encoder to emphasize significant features in logs for effective anomaly detection.  RTF maps unseen logs to their nearest neighbors among labeled logs, enabling adaptation to evolving logs for anomaly detection.

%Unsupervised tools are proposed to address data labeling costs (C1).

%Common methods used in unsupervised tools that eliminate data labeling costs (C2) include, e.g.,  PCA \cite{xu2009detecting}, clustering-based methods \cite{dani2017k}, isolation forest \cite{nyyssola2024speed}. 

\begin{table*}[htb]
\small
  \caption{Cross-system anomaly detection related works}
  \label{tab:CrossSys_related_work}
  \begin{tabular}{l p{3.1cm} p{1.3cm} p{3.3cm} p{6cm}}
    \toprule
    Tool & Method & AD level$^a$ & Training and evaluation context & Training data\\
    \midrule
    MetaLog \cite{zhang2024metalog}& Attention GRU + a meta-learning algorithm & Session & BGL→HDFS; \newline HDFS→BGL & Source system: \newline sufficient labeled log sessions \newline Target system: \newline 30\%/10\% normal sessions, \newline 1\% anomalous sessions. \\ \midrule

    LogTransfer \cite{chen2020logtransfer} & 2 fully connected LSTM networks %(one for the source system, another for the target system) 
    & Session &  HDFS →Hadoop; \newline
    SystemA/C→SystemB & Source system: \newline
    normal and anomalous sessions from
    HDFS (3,725,203), System~A (2,345,646), System C (525,427) respectively.
    \newline  Target system: 
    \newline 200 anomalous log sessions
    \\ \midrule
    
    %HDFS (3,725,203), System A (2,345,646), and System C (525,427) normal and anomalous sessions, respectively.
    
    % \newline 
    % 3,725,203, 2,345,646, 525,427 normal and anomalous sessions respectively from HDFS, System A, System C
    % - HDFS  3,725,203 sessions (both normal and anomalous)  \newline 
    % - System A 2,345,646 sessions
    % \newline 
    % - System A 525,427 \newline

    LogTAD \cite{han2021unsupervised} & Domain adversarial \newline LSTM & Event & BGL→Thunderbird; \newline Thunderbird→BGL  & Source system: \newline 100,000 normal sequences, \newline 100 anomalous sequences. \newline Target system: \newline 1000 normal sequences, \newline 10 anomalous sequences\\ \midrule

    OneLog \cite{hashemi2024onelog} & Character-level CNN  & Event & Liberty→Spirit$^b$; \newline Spirit→Liberty$^b$  & Source system: \newline All labeled log events from the system \newline Target system:\newline No labeled log  events\\

  \bottomrule
\end{tabular}
\scriptsize
\begin{flushleft} 
$^a$AD level = Anomaly Detection level: log-event level (Event) or log-session level (Session).\newline
$^b$For OneLog, we only report effective experiments: transfer learning from Liberty to Spirit and vice versa. 
\end{flushleft}
\end{table*}

In response to C1-C3, four studies propose transfer learning-based tools. Table \ref{tab:CrossSys_related_work} summarizes these tools.  These tools train on a source system with sufficient labeled logs and then adapt to a target system with little or no labeled logs. To create log representations, MetaLog, LogTransfer, and LogTAD parse raw logs into templates and use static methods (GloVe/Word2Vec) to generate word embeddings for these templates, while OneLog is parserless and utilizes a character-based statistical dictionary to produce character-level embeddings for log events. MetaLog, LogTransfer, and LogTAD employ sequence models, LSTM and Gated Recurrent Unit (GRU), while OneLog uses a convolutional neural network (CNN), to learn log sequences represented by their log embeddings to do anomaly detection. However, these tools fall short in fully addressing C1-C3, along with our study aim (Section \ref{sec:introduction}) and log data characteristics in our dataset (Section~\ref{sec:empirical_study}):

\begin{itemize}
    \item \textit{Model specificity restricts these tools to one-to-one transfer learning:} MetaLog, LogTransfer, and LogTAD require training on log data from both source and target systems. They are designed only for one-to-one transfer learning: from a single source system to a single target system. Specifically, MetaLog uses log data from both source and target systems to learn robust initial parameters that can adapt to the target system’s characteristics. LogTransfer and LogTAD utilize the shared components of models between the source and target systems to facilitate transferability. 
    Hence, all three tools must be retrained separately for each new pair of source and target systems, limiting their adaptability (C3) and offering minimal reduction in data labeling costs (C1) for organizations that operate more than two systems. In contrast, OneLog requires training only on the source system. However, its performance heavily depends on the relevance of the log data between the source and target systems \cite{hashemi2024onelog}. OneLog was trained individually on BGL, Thunderbird, Liberty, and Spirit, then evaluated on unseen systems during training. Training on Liberty and testing on Spirit, and vice versa, yielded high F1-Scores, while other experiments resulted in zero F1-Scores.  Thus, though OneLog is not designed for one-to-one transfer learning, it only works on some cases. It lacks robustness across different systems to address C3.
    
    %OneLog was trained individually on BGL, Thunderbird, Liberty, and Spirit, and then evaluated on systems that were not used during training.  The experimental results show that training OneLog on Liberty and evaluating it on Spirit, and vice versa, achieved high F1-Scores, while other experiments resulted in zero F1-Scores. 

    %The experimental results show that training on Liberty and evaluating on Spirit, and vice versa, achieved high F1-scores, while other combinations resulted in zero F1-scores. Thus, although OneLog is not specifically designed for one-to-one transfer learning, its final results demonstrate effectiveness only in such scenarios. It lacks robustness across different systems, and therefore, cannot fully address C3.

   % need to be trained using log data from both the source and target systems, which means it cannot be adapted to other new target systems without prior training on their log data. LogTransfer and LogTAD utilize the shared components of sequence models between the source and target systems to facilitate transferability. Thus, transfer learning with these tools is therefore limited to a one-to-one process—from one source system to a single target system, and vice versa.  Moving to a new target system necessitates retraining, which fails to meet the needs of organizations running multiple software systems (C3). 
    
    \item \textit{Static embeddings are limited for log-event level anomaly detection:} Our study focuses on log-event level anomaly detection, requiring precise interpretation of each log event while considering the specific context in which it is generated. Given the evolving nature (C2) and  the diverse, varied content (A3) of logs in software systems, static embeddings face limitations because they provide generalized, context-independent word/character representations in logs \cite{le2021log}. As logs continually evolve, static embeddings may struggle to adapt, making it difficult to capture the specific meanings and contextual details of individual log event. Moreover, static embeddings cannot handle OOV words, which frequently occur in software systems. For instance, system-specific terminologies, e.g., the terms used to indicate log sources (examples are given in Section \ref{sec:empirical_study}), often become OOV for static embeddings; referring to the log characteristic A2 in our observed distributed systems, OOV system-specific terms related to log sources can risk losing critical contextual information necessary for detecting anomalous log events tied to specific system components. 
    
    %such OOV system-specific terminologies related to log sources, ignoring them can obscure the context necessary for identifying anomalies tied to specific system components.

    %This level of detection requires precise interpretation of each log entry, where context and emergent vocabulary are crucial for identifying deviations from normal behavior.  Given the evolving nature (C2) and the diverse, dynamic content (A3) of logs in software systems, static embeddings face limitations because they provide generalized, context-independent word representations. As logs continually change in content and structure, static embeddings may fail to adapt to these changes, making it difficult to capture the specific meanings of individual log entries with contextual details. 
    
    %Given the evolving nature (C2) and diverse, varied content characteristics (A3) of logs in software systems, static embeddings struggle because they offer only a generalized, context-independent representation of words. They may fail to capture the specific meanings and implications of terms as used in individual log entries. Moreover, static embeddings can not address OOV (Out-Of-Vocabulary) words, which are frequent in dynamic software systems. For example, system-specific terminology used to indicate log sources often becomes OOV for static embeddings, further hindering their effectiveness in evolving environments. While "A2.Log source difference" (identified in Secition) chatecter is important to consider when we detect anomalies for log entires, why? 
\end{itemize}

Our CroSysLog builds upon existing research and provides novel insights. CroSysLog employs NeuralLog's neural representation approach \cite{le2021log} with the pre-trained model BERT  and subword tokenization to extract the semantics of logs, enabling a more nuanced understanding of log events in dynamic systems. CroSysLog extends prior transfer learning tools by going beyond one-to-one transfer learning. Prior tools are trained on both target and source systems to learn robust initial parameters or shared components of models, or rely on similar log data between both systems for transfer learning. In contrast, CroSysLog is trained using labeled log events from multiple source systems through a meta-learning approach, allowing it to learn robust model parameters that can efficiently adapt to new target systems for effective anomaly detection. CroSysLog emphasizes adaptability across diverse target systems and faster adaptation to new, unseen logs within these systems, making it more flexible and scalable for real-world applications.

%Three studies have developed AIOps tools to observe cross-system anomaly detection. Table \ref{tab:CrossSys_related_work} summarizes these tools. These tools perform log parsing and utilize static methods (GloVe or Word2Vec) to generate log embeddings. They employ sequence models (LSTM, GRU) to learn log sequences represented by these log embeddings. LogTransfer and LogTAD utilize the shared components of sequence models (GRU and LSTM) between the source and target systems to facilitate transferability. Their efficacy, however, depends on the relevance and compatibility of log data between the source and target system. Conversely, MetaLog employs the meta-learning algorithm, leveraging its ability to learn from the source system and then adapt to the target system. 

%\subsection{Our motivation}

\section{Methodology}

\begin{figure*}[ht]
\centering
\includegraphics[width=1\textwidth]{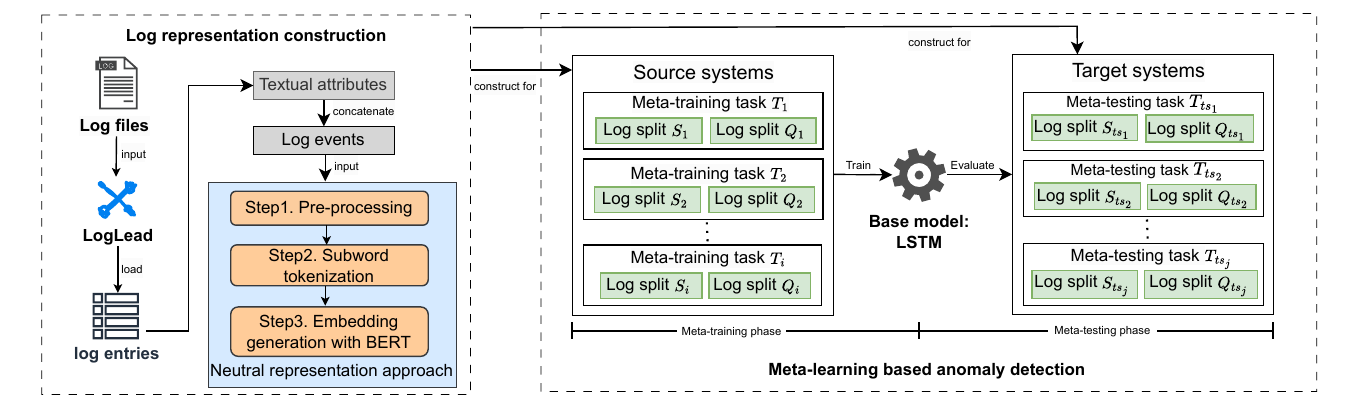}
\caption{CroSysLog overview}
\label{fig:framework_overview}
\vspace{-10pt}
\end{figure*}

 Figure \ref{fig:framework_overview} shows an overview of CroSysLog. CroSysLog consists of two components: (1) Log representation construction; (2) Meta-learning based log-event level anomaly detection. The design details on two components  of CroSysLog are described in Section \ref{sec:log_representation} and \ref{sec:meta-traing_design} respectively. We briefly introduce the workflow of CroSysLog below.   

\textit{Workflow of CroSysLog:}  CroSysLog trains a base model LSTM using a meta-learning approach \cite{finn2017model} to do log-event level anomaly detection for software systems. In the meta-training phase, CroSysLog is trained on meta-training tasks sampled from source systems. In the meta-testing phase, CroSysLog is evaluated on meta-testing tasks sampled from target systems. A meta-training/meta-testing task here refers to a training episode in meta-learning \cite{finn2017model}, denoting a single iteration where the base model is trained/evaluated on source/target system. The number of meta-training tasks and meta-testing tasks are not necessarily equal; typically, there are more meta-testing tasks to ensure comprehensive evaluation \cite{finn2017model}. Log splits in meta-training/meta-testing tasks come from the respective source/target systems \cite{finn2017model}. The representations of log events in these log splits are generated by our log representation construction approach.

\subsection{Log representation construction approach}
\label{sec:log_representation}
%Many prior studies use the common approach for creating log representations:  use a log parser (e.g., Drain, ) to extract log templates from log messages, construct log sequences using log templates, and then apply machine learning models on log sequences to detect anomalies.  

For a source/target system, we use a log loading and processing tool, LogLead \cite{mantyla2024without},  to load and process log files, separating individual log entries and organizing them into dataframes based on key attributes, e.g., timestamps, severity levels, reporting component, log messages, and anomaly labels. Log entries within each system are then sorted in chronological order to reflect the real-world operational context, and any entries with null values are removed. We extract these textual attributes from log entries: reporting component, severity level, log message. We concatenate these attributes into a single entity, referred as ``log event''.

To represent log events, we use NeuralLog's neural representation approach (mentioned in Section \ref{sec:related_work}), which enables the direct transformation of raw log events into semantic representations without log parsing, while also effectively handling new, changing, OOV words in evolving logs. Log parsing involves converting raw log data into a structured format, known as log templates, by removing variable components and retaining constant parts \cite{zhang2023system}. Based on empirical studies \cite{le2021log,ott2021robust} on log data of real-world software systems, log parsing is prone to errors such as OOV and semantic misunderstanding issues, which can negatively impact the precision of log representations and the effectiveness of anomaly detection. As logs continue to evolve in software systems (regarding C2 in Section \ref{sec:introduction}), maintaining accurate log parsing rules becomes increasingly difficult \cite{le2021log,ott2021robust}. In contrast, the neural representation approach addresses these issues effectively using subword tokenization, which breaks down words into subwords, and uses the pre-trained model BERT to generate embeddings for each subword. %This allows for handling new, changing, or OVV words that appear in logs but were not present in the training data. 
Such subword embeddings ensure that the semantic nuances of each component are captured, thereby maintaining high accuracy in understanding log contents despite the evolving nature of logs.

We construct neural representations on log events in three steps. \textit{Step1.Pre-processing.} We preprocess log events by converting all uppercase characters to lowercase, substituting sensitive variables with standardized identifiers (e.g., replacing ``P00:1A:2B:**:**:**'' with ``mac address'', ``/home/user/docs/file123.txt'' with ``file path''), and removing any non-alphabetic characters. \textit{Step2.Subword tokenization.} WordPiece tokenization technique \cite{wu2016google} is used to generate a subword vocabulary for log events, breaking down each log event into subwords. \textit{Step3. Embedding generation with BERT.} For each log event, subwords generated from the last step are fed into the BERT base model \cite{bert2018}, a pre-trained transformer-based model with 12 encoder layers and 768 hidden dimensions. Each encoder layer of the model generates embeddings for each subword. For each log event, we utilize its embeddings from the final encoding layer of the BERT base model and calculate the mean as the neural representation for this log event.

%In contrast, the neural representation approach addresses these issues effectively using subword tokenization, which breaks down words into subwords, and uses the pre-trained model BERT to generate word embeddings for each subword. This allows for handling new, changing, or OVV words that appear in logs but were not present in the training data. This decomposition into subwords ensures that the semantic nuances of each component are captured, thereby maintaining high accuracy in understanding and processing log data despite the dynamic and evolving nature of logs.

%For instance, a novel error code like ``ErrorPageNotFound'' can be decomposed into 
%``Error'', ``Page'', and ``Not'', ``Found'', each of which can be recognized individually even if the entire string is not. This decomposition into subwords ensures that the semantic nuances of each component are captured, thereby maintaining high accuracy in understanding and processing log data despite the dynamic and evolving nature of logs. 

\subsection{Meta-learning based anomaly detection} 
We use a meta-learning algorithm, MAML \cite{finn2017model}, to train a base model, LSTM, for log-event level anomaly detection. The following subsections describe the base model, and meta-training and meta-testing phases in detail.

%This section explains how CroSysLog utilizes its LSTM-based model for anomaly detection and describes its integration with the MAML algorithm \cite{finn2017model} for cross-system adaptability.

\label{sec:meta-traing_design}
%For clarity, we define meta-learning terminologies used in our study. ``Meta-training tasks" refer to the set of training tasks to train our CroSysLog. "Meta-testing tasks" are evaluation tasks to evaluate the performance of our CroSysLog. In the context of cross-system learning, we use the term "source systems" to denote the software systems from which we sample meta-training tasks for training our CroSysLog, and the term "target systems" refers to any software systems from which we sample meta-testing tasks for evaluating our CroSysLog. Since CroSysLog is based on the MAML algorithm that enables multi-task learning, meta-training tasks can be sampled from more than one source system. Similarly, meta-testing tasks can also be sampled from multiple target systems because, after training, our CroSysLog is designed to adapt to any anomaly tasks across any software system.

\subsubsection{Base model for anomaly detection} We use LSTM as the base model due to its capability to capture long-term dependencies \cite{hochreiter1997long}. In our datasets,  logs are generated by multiple components in distributed systems, resulting in logs that do not necessarily follow a strict sequential order. Anomalies may arise from log events that occur far apart in time or across different components. LSTM's architecture is well-suited for detecting whether a log event is anomalous or not by analyzing its context, where distant log events may provide important information. This ability to capture dependencies across non-adjacent log events allows LSTM to identify anomalies that would otherwise go unnoticed in isolated log events.

%to do log event-based anomaly detection (i.e., detect anomalies in log lines). Following previous studies that commonly use sequence-based models, we choose LSTM due to its proficiency in handling sequence data, which is essential given the unique nature of log data in our datasets. In our datasets, logs originate from distributed systems where different components generate logs. This often results in non-sequential data that may appear to lack continuous sequential order. LSTM's capability to remember long-term dependencies makes it highly suitable for such scenarios as it can effectively link distant log entries, capturing relevant features for anomaly detection.

For a specific system, given a log split $L$ that contains consecutive log events in chronological order, $L_i$ refers to the $i$-th log event in this log split. $L$ is represented by log representation $V$, where each $L_i$ is represented by $V_i$. $V$ is generated using our log representation approach in Section~\ref{sec:log_representation}. We apply a fixed-window method to split $L$ into non-overlapping windows of logs, each window is defined as: 
\begin{equation}
W_j = \{V_i, V_{i+1}, \dots, V_{i+k-1}\}
\end{equation}
here, $W_j$ contains $k$ consecutive log events starting from  $V_i$  within $V$, and $k$ is the window size. After processing window $W_j$, the next window $W_{j+1}$ starts from $V_{i+k}$. The equation for processing each $V_i$ within a window $W_j$ is:

\begin{equation}
\label{eq:LSTM}
h_i = \text{LSTM}(V_i, h_{i-1})
\end{equation}
where $h_i$ is the hidden state for $V_i$, updated based on the previous hidden state $h_{i-1}$ and the current log event $V_i$, effectively capturing dependencies among log events within a window $W_j$. We pass $h_i$ through a fully connected layer, which maps it to an output vector corresponding to the prediction classes, anomalous and normal. Subsequently, the softmax function is applied to this vector to generate a probability distribution over two classes for each log event.

We apply the fixed-window method instead of processing the entire $L$ at once to ensure our base model effectively detects context-dependent anomalous log events that result from interactions with nearby log events. Processing log events in manageable windows reduces computational complexity \cite{xie2024logsd} by allowing the model to learn meaningful patterns without being overwhelmed by distant, irrelevant data.

\subsubsection{Meta-training phase} We randomly sample meta-training tasks from source systems, collectively denoted as~$T$. Each task is unique and denoted as $T_i$, consisting of a pair $(S_i, Q_i)$, where $S_i$ and $Q_i$ are distinct log splits of a source system. In each task $T_i$, $S_i$ serves as the support set, providing a few labeled log events for task-specific adaptation for a source system; the query set $Q_i$ evaluates the base model's performance on new, unseen log events after adaptation for this source system. Thus, $T$ is a tuple $(S, Q)$, $S$ and $Q$ represent all support and query sets for $T$. 

% Log representations for logs in $S$ and $Q$ are generated using our log representation construction approach in Section \ref{sec:log_representation}.

\begin{algorithm}
\caption{Meta-training phase operations}
\label{ag:meta-traing}
\small 
\begin{algorithmic}[1]
\Require $T (S, Q)$, base model LSTM
\Require task-level learning rate $\alpha$, meta-learning rate $\beta$
\State Randomly initialize base model LSTM with parameters $\theta$
\While{not done}
    \ForAll{$T_i = (S_i, Q_i)$ in $T$}
        \State Compute $\nabla_{\theta} \mathcal{L}_{S_i} (\theta)$ using $S_i$ and loss function $\mathcal{L}_{S_i}$

        \State Compute task-level adapted parameters $\theta'_i$ on $S_i$ with 
        
        \Statex \hspace{\algorithmicindent}\hspace{\algorithmicindent} gradient descent updates, each update is:
        
        \Statex \hspace{\algorithmicindent}\hspace{\algorithmicindent} $\theta'_i = \theta - \alpha \nabla_{\theta} \mathcal{L}_{S_i}(\theta)$

        \State Evaluate task-level adapted parameters $\theta'_i$ on $Q_i$ to \Statex \hspace{\algorithmicindent}\hspace{\algorithmicindent} compute $\mathcal{L}_{Q_i} (\theta'_i)$

    \EndFor
    \State Update $\theta$ by applying gradient descent updates over all tasks \Statex \hspace{\algorithmicindent}~$T$, each gradient decent update is: 
    \Statex \hspace{\algorithmicindent} $\theta \leftarrow \theta - \beta \nabla_{\theta} \sum_{i=1}^n \mathcal{L}_{Q_i}(\theta'_i)$
\EndWhile

\noindent\Return an optimized base model LSTM  with parameters $\theta^*$
\end{algorithmic}
\end{algorithm}

Our meta-training phase is conducted in multiple steps, as outlined in Algorithm \ref{ag:meta-traing}. First, we randomly initialize our base model LSTM with parameters $\theta$. Second, we adapt our base model LSTM to each task $T_i$, using this task's $S_i$ with the current model parameters $\theta$. This process generates adapted parameters $\theta'_i$ for $T_i$ through few gradient descent updates. Each update is formulated as:

\begin{equation}
\label{eq:innerUpdate}
\theta'_i = \theta - \alpha \nabla_{\theta} \mathcal{L}_{S_i}(\theta)
\vspace{-3pt}
\end{equation}
where $\alpha$ is the task-level learning rate and $\mathcal{L}_{S_i}(\theta)$ is the loss on $S_i$ for $T_i$. After task-specific adaptation, we optimize our base model LSTM's parameters $\theta$ across all tasks $T$ to ensure robust adaptation to all tasks $T$. This is achieved through few gradient descent updates, each formulated as: 

\begin{equation}
\label{eq
}
\theta = \theta - \beta \nabla_{\theta} \sum_{i=1}^n \mathcal{L}_{Q_i}(\theta'_i)
\end{equation}
where $\beta$ is the meta-learning rate, and $\mathcal{L}_{Q_i}(\theta'_i)$ is the loss on $Q_i$ for $T_i$. This update step refines the model parameters $\theta$ by aggregating insights from the performance of $\theta'_i$ on its corresponding $Q_i$ of $T_i$. The objective is to minimize the overall loss for all meta-training tasks $T$, ensuring that the updated parameters generalize well across all tasks. The optimization problem can be formulated as follows: 

\begin{equation}
\label{eq}
\min_{\theta} \sum_{i=1}^n \mathcal{L}_{Q_i}(\theta - \beta \nabla_{\theta} \mathcal{L}_{Q_i}(\theta'_i))
\vspace{-3pt}
\end{equation}

Following prior MAML studies \cite{finn2017model,ye2021train}, we apply the first-order approximation technique to simplify the above update process by avoiding the computation of second-order derivatives. This approximation is crucial for efficiently scaling the meta-training process while balancing performance and computational costs. In the end, we obtain the optimal parameters $\theta^*$ through this streamlined optimization process. We then update our base model LSTM  using $\theta^*$, resulting in an optimized version of the model. 

%This optimized model is better equipped to quickly adapt to new, unseen anomaly detection tasks from other software systems. 

\subsection{Meta-testing phase} In this phase,  we use the optimized LSTM (with parameters~$\theta^*$) on meta-testing tasks, denoted as $T_{ts}$, which are sampled from target systems, to evaluate our framework. Each meta-testing task $T_{ts_j}$ is unique and follows the same structure as each meta-training task $T_{i}$: $T_{ts_j}$ consists of a pair $(S_{ts_j}, Q_{ts_j})$, which are distinct log splits of a target system. Thus,  $T_{ts}$ = $(S_{ts}, Q_{ts})$, where  $S_{ts}$ and $Q_{ts}$ refer to support and query sets of all meta-testing tasks respectively. For each meta-testing task $T_{ts_j}$ of a target system, our optimized LSTM is fine-tuned on labeled log events of  $S_{ts_j}$, to rapidly adapt its parameters to this task; after fine-tuning, we apply our task-specific LSTM on $Q_{ts_j}$ for evaluating the performance of our CroSysLog in log-event level anomaly detection.

\section{Experiment design}
\subsection{Research questions}
%\textbf{}To evaluate our framework, CroSysLog, 
To evaluate CroSysLog, we define these research questions: 
\begin{itemize}
    \item \textbf{RQ1: Anomaly detection effectiveness.} How effective is CroSysLog, once trained, in anomaly detection on target systems compared with baseline approaches?
    
    \item \textbf{RQ2: Efficiency.} How efficient is CroSysLog in model training and anomaly detection compared with baseline approaches?
\end{itemize}

\subsection{Datasets} We use logs from four large-scale distributed supercomputing systems:
BGL, Thunderbird,  Liberty, and Spirit, with open datasets. The detailed description for each dataset and log data characteristics are provided in Section~\ref{sec:empirical_study}.

\subsection{Training and evaluation}
\label{sec:training_and_evaluation}
We utilize Liberty and BGL as source systems to train CroSysLog, and Thunderbird and Spirit as target systems to evaluate its performance. Our choice is made to account for differences in the proportion of anomalous log events and log collection duration of these systems, see Table \ref{tab:stat_datasets}. BGL and Thunderbird have a lower proportion of anomalous log events, while Liberty and Spirit exhibit a higher proportion. BGL's and Thunderbird's logs were collected within a year, whereas Liberty's and Spirit's logs span two years, suggesting that Liberty and Spirit datasets may include more evolving logs. Training CroSysLog on Liberty and BGL ensures it learns from diverse, varied log data from these source systems, while evaluating CroSysLog on Thunderbird and Spirit allows for assessing its performance on these target systems with varying anomaly distributions and log collection duration, providing a balanced and realistic evaluation context. By using diverse, varied logs from multiple source systems, we aim to enhance CroSysLog's robustness and adaptability efficiency, leveraging the MAML algorithm's effectiveness in learning from diverse data to enable rapid adaptation to new, unseen data\cite{finn2017model}. We conduct ablation studies (Section \ref{sec:ablation_study}) to demonstrate that the performance of CroSysLog is not driven by the choice of the current source systems. %This is particularly important due to the clustered distribution of normal/anomalous log entries observed in each system (Section \ref{sec:empirical_study}).

%specially in light of the clustered distribution of normal/abnormal log entries within each system (Section \ref{sec:empirical_study}). Using diverse, varied logs from multiple source systems can enhance CroSysLog's adaptability efficiency, as it is based on the MAML algorithm, known for its effectiveness in learning from varied data and enabling rapid adaptation to new, unseen data \cite{finn2017model}.

We randomly select log splits from source systems, Liberty and BGL, to build meta-training tasks for training our CroSysLog. Specifically, we select two log splits from each source system. Each log split consists of 10,000 consecutive log events,  of which 0.01-0.5\% are anomalous. Each log split is unique and does not overlap with others. We set this broad range for the proportion of anomalous log events to account for log distribution variance and the clustered nature of normal/anomalous log events in our datasets (detailed in Section \ref{sec:empirical_study}), in order to maximize randomness for obtaining varied log splits from each source system. We build each meta-training task $T_i$ for each source system: we use one log split as the support set $S_i$ and another as the query set $Q_i$. In total, we build 2 meta-training tasks as $T$, one per source system.

We randomly select log splits from target systems, Thunderbird and Spirit, to build meta-testing tasks for evaluating our CroSysLog. These splits were widely distributed across a one/two-year span
of each system’s log collection duration, capturing the
evolving nature of the logs in each system. Specifically, for each target system, we build 20 meta-testing tasks $T_{ts}$=($S_{ts}$, $Q_{ts}$). Each task $T_{ts_j}$ is constructed as follows: a random log split with 2,000 consecutive log events is used as the support set $S_{ts_j}$, and another random log split with 10,000 consecutive log events is used as the query set $Q_{ts_j}$; each log split is unique and does not overlap with others. For Thunderbird, each $S_{ts_j}$ contains 0.2-0.5\% anomalous log events, while each $Q_{ts_j}$ contains 0.01-0.5\% anomalous log events.  For Spirit, each $S_{ts_j}$ contains 0.2-0.7\% anomalous logs, while each $Q_{ts_j}$ contains 0.01-0.7\% anomalous log events.

For both target systems, each support set $S_{ts_j}$ is a log split with only 2000 consecutive log events, because CroSysLog just requests a few labeled log events with a certain proportion of anomalies to rapidly adapt to the target system, leveraging the MAML algorithm for efficient adaptation. We ensure that each $S_{ts_j}$ includes a sufficient proportion of anomalies, enabling CroSysLog to effectively learn the distinguishing features between anomalous and normal logs, thereby adapting to the target system's context. In contrast, for each target system, each query set $Q_{ts_j}$ has 10,000 consecutive log events, as this larger size provides a more comprehensive evaluation after adaptation with the support set $S_{ts_j}$; Each $Q_{ts_j}$ has a broader range of anomalous log proportions (0.01-0.5\% for Thunderbird, 0.01-0.7\% for Spirit) to ensure involving a sufficient variety of log splits for a comprehensive evaluation of CroSysLog. This design also considers the characteristic A1 of our datasets (Section \ref{sec:empirical_study}): Spirit has a high prevalence of anomalous log events while Thunderbird contains a large proportion of normal log events, and normal/anomalous log events may cluster in each system. For each target system, log splits used for building query sets $Q_{ts}$ are widely distributed across the log collection duration years, considering the large size of both Thunderbird and Spirit datasets, which were collected within a year/two years, see Table \ref{tab:stat_datasets}. This strategy ensures that evolving logs within target systems are included, in response to C2 (detailed in Section \ref{sec:introduction}) in the evaluation of CroSysLog. 

%we select 40 log sequences: 20 log sequences each containing 2,000 log lines, and the remaining 20 log sequences each consisting of 10,000 log lines, all arranged chronologically.  Each log sequence is unique and does not overlap with others. For Thunderbird, each log sequence with 2000 log lines contains 0.2-0.5\% anomalous log lines, while each log sequence with 10,000 log lines contains 0.001-0.5\% anomalous log lines. For SPIRIT, each log sequence with 2000 log lines contains 0.2-0.7\% anomalous log lines, while each log sequences with 10,000 log lines contain 0.01-0.7\% anomalous log lines.

%Using selected 40 log sequences for each target system, we build 20 meta-testing tasks, each task $T_{test}$ is constructed as follows: we use a random log sequence with 2000 log lines as the support set $S_{test}$ and the random log sequence with 10,000 log lines as the query set $Q_{test}$. 

Using the established meta-training and meta-testing tasks for source and target systems, we design our experiments: 

\begin{itemize}
    \item \textbf{E1 (Liberty+BGL→Thunderbird).} We train CroSysLog on 2 meta-training tasks (one from Liberty and another from BGL) and then evaluate it on Thunderbird’s 20 meta-testing tasks. 
    \item \textbf{E2 (Liberty+BGL→Spirit).} We train CroSysLog on 2 meta-training tasks (one from Liberty and another from BGL) and then evaluate it on Spirit’s 20 meta-testing tasks. 
\end{itemize}

\subsection{Implementation details} We conduct experiments on a Linux server with a 32-core CPU and an
NVIDIA Ampere A100 GPU with 40 GB of memory, utilizing Python 3.10.6. We train our framework using the AdamW optimizer. Further details regarding hyperparameter settings are provided in our replication package. 

%We will publish our replication package upon acceptance. The anonymous link for review: \cite{}

\subsection{Baselines}
\label{sec:baselines}
We consider the related work in Section \ref{sec:related_work} and build two types of baselines: transfer learning, and supervised. Unsupervised baselines are excluded because they lack adaptability. We tested some of them (e.g., k-means, k-nearest neighbors, isolation forest) on meta-testing tasks $T_{ts}$ of each target system, training on these tasks' support sets $S_{ts}$ and evaluating on these tasks' query sets $Q_{ts}$ to mimic the adaptation phase of our CroSysLog to the target system. They got F1-Scores mostly below 40\%, making them unsuitable for further comparison. %The potential reason could be that they rely on each individual log event for anomaly detection without capturing its contextual information.
Training and evaluation contexts of our baselines are detailed in Table \ref{tab:E1_results}-\ref{tab:E2_results}. %We described our baselines below.

%consistently achieve low F1-scores (mostly < 60\%) when using the same amount of log entries from the target system as our framework uses to adapt to this target system context. Unsupervised baselines are excluded because they lack learning capabilities and they achieve low F1 (mostly < 30\%) when training using the same amount of logs from the target system as our framework to adapt to the target system on our meta+testing tasks of each target systemUnsupervised baselines are excluded because they lack transfer learning capabilities and rely on thresholds for anomaly detection, making them unsuitable for our meta-testing tasks, which contain a broad range of anomalous log proportions. 

We use MetaLog \cite{zhang2024metalog}, LogTransfer \cite{chen2020logtransfer}, and LogTAD \cite{han2021unsupervised} to build transfer learning  baselines.  We omit OneLog \cite{hashemi2024onelog} because, in most of its experiments, the F1-Score is zero in cross-system contexts, indicating poor robustness (detailed in Section \ref{sec:related_work}). As described in Section \ref{sec:related_work}, MetaLog, LogTransfer, and LogTAD require training on log data from both source and target systems, performing one-to-one transfer learning: from a single source system to a single target system. We design the training and evaluation context of these baselines to maintain their original one-to-one transfer learning design while ensuring consistency with CroSysLog. In each experiment, we train each of these baselines using each source system's meta-training task $T_i$ (utilizing this task's both the support set $S_i$ and the query set $Q_i$) seperatly, combined with $S_{ts_j}$ from a random meta-testing task $T_{ts_j}$ of the target system; we then evaluate each baseline's performance for anomaly detection on query sets $Q_{ts}$ of the target system's 20 meta-testing tasks, just as CroSysLog does. %Training contexts of these baselines are detailed in Table \ref{tab:E1_results} \& \ref{tab:E2_results}.

We build supervised baselines using LogRobust \cite{zhang2019robust}, NeuralLog \cite{le2021log}, and RTF \cite{ott2021robust}. As both LogRobust and SwissLog use Bi-LSTM to model log sequences for anomaly detection (detailed in Section \ref{sec:related_work}), we included LogRobust as a baseline but excluded SwissLog to avoid duplication. Since these supervised baselines do not support transfer learning, to align their training and evaluation contexts with our CroSysLog, in each experiment, we train them on each target system using the support set $S_{ts_j}$ of each meta-testing task $T_{ts_j}$ from that system, and evaluate them on the corresponding query set $Q_{ts_j}$ from each meta-testing task for that system. The rationale behind this design is to evaluate whether these baselines can achieve comparable effectiveness and efficiency (RQ1 \& RQ2) when provided with the same number of labeled log events that our CroSysLog uses to adapt the target system. If they perform similarly, it may indicate that transfer learning offers no additional benefit in this context. 

We develop all baselines for log-event level anomaly detection, including those (MetaLog, LogTransfer, LogRobust, NeuralLog, RTF) originally designed for log-session level anomaly detection by modifying their models to fit our study context. Besides, we adopt the neural representation approach (Section \ref{sec:log_representation}) in all baselines to construct log representations. As described in Section \ref{sec:related_work}, original models of many baseline approaches use log parsing and static models to construct log representations. Empirical studies \cite{le2021log,ott2021robust,wang2024few} have shown that the neural representation approach is more effective in capturing the semantic meaning of evolving logs. Our modification ensures that any observed differences in effectiveness between our CroSysLog and the baselines are not solely due to the neural representation approach used in CroSysLog. 

\subsection{Metrics}
To measure the effectiveness of anomaly detection of our CroSysLog and baselines (RQ1), we use Precision, Recall and F1-Score as metrics. These metrics are calculated based on True Positives (TP), False Positives (FP), and False Negatives (FN) as follows: Precision = \(\frac{TP}{TP + FP}\), Recall = \(\frac{TP}{TP + FN}\), and F1-Score = \(\frac{2 \cdot \text{Precision} \cdot \text{Recall}}{\text{Precision} + \text{Recall}}\). These metrics are widely used for anomaly detection because they together offer a comprehensive evaluation. To measure the efficiency of our CroSysLog and baselines in model
training and anomaly detection (RQ2), we calculate training and testing time.%Precision measures the proportion of correctly detected anomalous log entries among all log entries that are detected as anomalous.
%Recall measures the proportion of correctly detected anomalous log entries among all actual anomalous log entries. F1-Score is the harmonic mean of Precision and Recall giving an overall measure. 

%As we use binary F1 the models are not rewarded by correctly predicting the majority class (non-anomaly) 

%These metrics are calculated based on True Positives (TP), False Positives (FP), and False Negatives (FN) as follows: Precision = \(\frac{TP}{TP + FP}\), Recall = \(\frac{TP}{TP + FN}\), and F1 = \(\frac{2 \cdot \text{Precision} \cdot \text{Recall}}{\text{Precision} + \text{Recall}}\). These metrics are widely used for anomaly detection because they together offer a comprehensive evaluation. Precision measures the proportion of correctly detected anomalous log entries among all log entries that are detected as anomalous.
%Recall measures the proportion of correctly detected anomalous log entries among all actual anomalous log entries. F1-Score is the harmonic mean of Precision and Recall giving an overall measure. 

\section{Experiment results and analysis}
%\subsection{RQ1.Effectiveness and efficiency.}

\begin{table*}[!ht]
%\caption[]{\small TABLE \thetable. Liberty+BGL→Thunderbird: training and testing time (in seconds)}
%\captionsetup{justification=centering, labelsep=colon, font=normalfont} % Local to this table % Manual captio

\caption{\textnormal{Evaluation of CroSysLog and baselines on Thunderbird (using its 20 meta-testing tasks' query sets) in E1.}}
\label{tab:E1_results}
\centering
\begin{tabular}{l llll}
\toprule
Method & Training and evaluation context & Precision & Recall & F1-Score \\ \midrule
Our CroSysLog & Liberty+BGL (2 meta-training tasks, one per each system)→Thunderbird & 96.99 & 98.17 & 97.55  \\ \hline
%Our CroSysLog & LIBERTY+SPIRIT→Thunderbird & 90.38 & 91.73 & 90.84 \\ \hline

%MetaLog & LIBERTY+BGL+TB (only $Q_{test}$)→TB & & &  \\
%MetaLog & LIBERTY+TB (only $Q_{test}$)→TB & 31.13 & 93.97 & 45.84 \\
%MetaLog & BGL+TB (only $Q_{test}$)→TB & 79.23 & 99.97 & 87.47 \\ \hline

MetaLog (a)&  Liberty (1 meta-training task: $S_i$+$Q_i$) + Thunderbird (a random $S_{ts_j}$)→Thunderbird & 31.13 & 93.97 & 45.84 \\
MetaLog (b)& BGL (1 meta-training task: $S_i$+$Q_i$) + Thunderbird (a random $S_{ts_j}$)→Thunderbird & 26.85 & 86.54 & 38.50\\ \hline

LogTransfer (a)  & Liberty (1 meta-training task: $S_i$+$Q_i$) + Thunderbird (a random $S_{ts_j}$)→Thunderbird & 99.71 %27epoch 
& 96.10 & 97.88 \\
LogTransfer (b) & BGL (1 meta-training task: $S_i$+$Q_i$) + Thunderbird (a random $S_{ts_j}$)→Thunderbird & 99.82 & 93.17 & 96.38 \\\hline

LogTAD (a) & Liberty (1 meta-training task: $S_i$+$Q_i$) +Thunderbird (a random $S_{ts_j}$)→Thunderbird  &16.49 & 90.91 & 27.73 \\  
LogTAD (b) & BGL (1 meta-training task: $S_i$+$Q_i$) +Thunderbird (a random $S_{ts_j}$)→Thunderbird & 33.53 & 98.64& 49.52  \\\hline

LogRobust & Thunderbird (each $S_{ts_j}$)→Thunderbird & 14.37& 100.0 & 24.75  \\ \hline
NeuralLog & Thunderbird (each $S_{ts_j}$)→Thunderbird & 85.00  &68.33 & 72.30  \\ \hline
RTF & Thunderbird (each $S_{ts_j}$)→Thunderbird & 54.69 & 99.95 & 70.69 \\ 
%SwissLog & & \\ \hline
\hline 
%LstmAE & Thunderbird (only normal logs)→Thunderbird & 59.68 & 55.89 & 46.21 \\\hline

%Isolation tree & & 0.1 & 0.02 & 0.04 \\ 

\end{tabular}
\vspace{-5pt}
\end{table*}

\begin{table*}[!ht]
\caption{\textnormal{Evaluation of CroSysLog and baselines on Spirit (using its 20 meta-testing tasks' query sets) in E2.}}

\label{tab:E2_results}
\centering
\begin{tabular}{l l lll}
\toprule
Method & Training and evaluation context & Precision & Recall & F1-Score \\ \midrule
Our CroSysLog & Liberty+BGL (2 meta-training tasks, one per each system)→Spirit & 98.38 & 99.99 & 99.17 \\ \hline
%Our & TB+BGL→SPIRIT & 99.66 & 99.99 & 99.82  \\ \hline
%MetaLog & LIBERTY+BGL+SPIRIT (only $Q_{test}$)→SPIRIT & & &  \\ 
MetaLog (c) & Liberty (1 meta-training task: $S_i$+$Q_i$) + Spirit (a random $Q_{ts_j}$)→Spirit & 67.31 & 98.65 & 79.25  \\
MetaLog (d) & BGL (1 meta-training task: $S_i$+$Q_i$) + Spirit (a random $Q_{ts_j}$)→Spirit &64.04 &99.99& 77.81 \\ \hline

LogTransfer (c) & Liberty (1 meta-training task: $S_i$+$Q_i$) + Spirit (a random $Q_{ts_j}$)→Spirit & 99.87 & 99.97 & 99.92 \\
LogTransfer (d) & BGL (1 meta-training task: $S_i$+$Q_i$) + Spirit (a random $S_{ts_j}$)→Spirit & 99.99 & 96.28 & 98.10 \\\hline

LogTAD (c) & Liberty (1 meta-training task: $S_i$+$Q_i$) + Spirit (a random $Q_{ts_j}$)→Spirit & 59.35& 51.08 & 47.72  \\
LogTAD (d) & BGL (1 meta-training task: $S_i$+$Q_i$) + Spirit (a random $Q_{ts_j}$)→Spirit & 60.48& 55.46& 48.10 \\\hline

LogRobust & Spirit (each $S_{ts_j}$)→Spirit & 41.22& 100.0 & 58.26   \\\hline
NeuralLog & Spirit (each $S_{ts_j}$)→Spirit & 49.31  &100.0 & 57.17  \\ \hline
RTF & Spirit (each $S_{ts_j}$)→Spirit & 14.79 & 56.48 & 25.25  \\
\hline
%LstmAE & SPIRIT (only normal logs)→SPIRIT & 49.96 & 40.53& 38.20  \\ \hline
\end{tabular}
\vspace{-5pt}
\end{table*}

\begin{table*}[!ht]
\caption{Traning and testing time (in seconds)}
\label{tab:model_efficiency}
\centering
%\caption{Baselines.}
\label{tab:baselines}
\begin{tabular}{lrr | lrr}
\toprule
E1 &  & & E2 &  &  \\ 
Method & Training  time$^a$ & Testing time$^b$ & Method &  Training  time & Testing time \\ \hline 
Our CroSysLog & 6.5419 %12epoch 
& 0.6328 & Our CroSysLog & 3.8692 %4epoch 
& 0.6414  \\ 
MetaLog (a) & 39.6874 %37 epoch 
&1.0728& MetaLog (c)  & 12.7600 %12 epoch 
& 1.0233\\
MetaLog (b) & 15.8640 %9epoch 
& 1.0505 & MetaLog (d) & 11.8289 & 1.2910\\
LogTransfer (a) & 53.1888& 1.6387& LogTransfer (c) & 21.2169 & 1.6707\\
LogTransfer (b) & 16.4155 & 1.4288  & LogTransfer (d) & 18.8828 & 1.4011 \\
LogTAD (a) & 110.3545 & 1.7622 & LogTAD (c) &  121.4121 & 1.7333 \\
LogTAD (b) & 115.1244 & 1.6868 & LogTAD (d) & 131.4512 & 1.8101 \\
LogRobust & 37.9406 %{22epoch}
& 1.6055 & LogRobust & 18.2693 & 1.5883 \\ 
NeuralLog & 17.5623 & 2.3485 & NeuralLog & 16.0592 & 2.2241 \\
RTF & 0.0913 & 1.2392 & RTF & 0.0551 & 0.8433 \\
%LstmAE & 15.3497 & 1.6971 & LstmAE & 18.9993 & 1.6706   \\
\bottomrule
\end{tabular}
\scriptsize
\begin{flushleft} $^a$Training time refers to the time used to train each method excluding the time spent on constructing log representations. This exclusion is made because, though the neural representation approach is applied for both CroSysLog and baselines (as described in Section \ref{sec:baselines}), differences in their training contexts (see Table \ref{tab:E1_results}-\ref{tab:E2_results}) could introduce time variations; however, we focus on comparing the model performance of each method without the influence of such variations.
\newline $^b$Testing time refers to the time each method is evaluated on query sets $Q_{ts}$ of 20 meta-testing tasks in each experiment.
\end{flushleft}
\vspace{-13pt}
\end{table*}

\subsection{RQ1: Anomaly detection effectiveness}
Table \ref{tab:E1_results}-\ref{tab:E2_results} present our experimental results on three metrics—Precision, Recall, F1-Score—for E1 and E2, respectively. Our CroSysLog achieves a high F1-Score in both experiments (E1: 97.55\%, E2: 99.17\%), indicating that, after training on varied logs from source systems (Liberty and BGL), it is highly effective at performing log-event level anomaly detection on different target systems (Thunderbird and Spirit) after using a few labeled log events of each system for adaption. E1 and E2 were trained using the same meta-training tasks from Liberty and BGL. This demonstrates that our CroSysLog is not limited to one-to-one transfer learning, but rather shows broader adaptability across multiple target systems after training. 

LogTransfer achieves similar F1-Scores to our CroSysLog in E1 and E2. This suggests that, our CroSysLog, which enables adaptability across multiple target systems after training, can get the same level of effectiveness in anomaly detection as LogTransfer, which only performs one-to-one transfer learning (from a single source system to a single target system) and requires training on log data from both systems. This confirms that our CroSysLog offers broader adaptability while maintaining comparable effectiveness for organizations that run many software systems. Additionally, our CroSysLog is much more effective than other transfer learning baselines, as evidenced by the fact that Metalog and LogTAD related baselines achieve F1-Scores below 80\% in E1-E2. 

Supervised baselines (LogRobust, NeuralLog, RTF) only achieve F1-Scores in the range of 24.75\% to 72.30\% in E1-E2, as shown in Table \ref{tab:E1_results}-\ref{tab:E2_results}. This indicates that, when these supervised baselines are trained on each target system, using the same amount of labeled log events as our CroSysLog uses for adapting to each target system, these supervised baselines fail to match CroSysLog's effectiveness. This highlights CroSysLog's advantage, which only needs a few labeled log events from the target system for adaptation and can then perform effective anomaly detection for this system.

\subsection{RQ2: Efficiency}
Table \ref{tab:model_efficiency} compares our CroSysLog's training time and testing time with baselines. CroSysLog is most efficient among all methods, with a training time of 6.5419 seconds and a testing time of 0.6328 seconds in E1, and a training time of 3.8692 seconds and a testing time of 0.6414 seconds in E2. RTF has the fastest training time among all methods, with 0.0913 seconds in E1 and 0.551 seconds in E2, but it requires approximately twice as much testing time in E1 and one-third more in E2 compared to CroSysLog. For other baselines, our framework requires about one-third of training time and uses less than 60\% of the testing time. 

The efficiency of our CroSysLog is attributed to the MAML algorithm. In CroSysLog, MAML trains the base model LSTM with a few gradient steps 
across meta-training tasks, significantly reducing the overall training time in each experiment; after training, the base model LSTM rapidly adapts to the target system with a few labeled log events also using just a few gradient descent steps. %(for our experiments: 12 steps in E1, 4 steps in E2)
CroSysLog requires less time also because baselines have more complex model mechanisms than our base model LSTM to perform anomaly detection. Specifically, MetaLog and LogRobust baselines incorporate the attention mechanisms to their sequence models (LSTM/GRU); NeuralLog uses a Transformer encoder embedded with a self-attention mechanism; LogTransfer baselines use two connected LSTM networks; RTF uses the nearest neighbor method, for each meta-testing task $T_{ts_j}$, it searches the labeled log events in this task's support set $S_{ts_j}$ to find the closest match for each log event in this task's query set $Q_{ts_j}$.

%RTF uses the nearest neighbor method, for each meta-testing task in each experiment, it searches the labeled log entries in this task's the support set to find the closest match for each log entry in this task's query set. 

%Training time here refers to the time used to train each method excluding the time spent on constructing log representations. This exclusion is made because, though the neural representation approach is applied for both CroSysLog and baselines (as described in Section \ref{sec:baselines}), differences in their training contexts (see Table \ref{tab:E1_results}-\ref{tab:E2_results}) could introduce time variations; however, Our goal is to focus on comparing model performance of each method without the influence of such variations. Testing time in Table \ref{tab:model_efficiency} refers to the time each method is evaluated on query sets $Q_{ts}$ of 20 meta-testing tasks in each experiment.

%\subsection{Discussion}
% Based on the above evaluation results, our CroSysLog demonstrates significant advantages in addressing practical challenges C1-C3, as discussed below: 
%\textit{Answers to our research questions:} based our evaluation results, we can conclude that our CroSysLog has more  effective  

%Our CroSysLog addresses data labeling costs (C1) by allowing training on sufficiently labeled logs from available open datasets, while requiring only a small number of labeled logs from new systems for adaptation.
\vspace{-2pt}
\subsection{Ablation study}
\label{sec:ablation_study}
Table \ref{tab:ablation_study} presents our ablation studies for each experiment, evaluating the contribution of CroSysLog's main components to its effectiveness. In these ablation studies, we utilize our already established meta-training tasks for source systems Liberty and BGL. For Thunderbird and Spirit, which are not used as source systems in E1-E2 before, we follow the same steps outlined in Section \ref{sec:training_and_evaluation} to prepare meta-training tasks for them, ensuring that their training contexts are consistent with those used in E1-E2. Our ablation studies are evaluated on 20 meta-testing tasks of the target system in each experiment. 

We investigate whether the effectiveness of our CroSysLog is solely driven by using logs from the current source systems, Liberty and BGL. For each experiment, we use different source systems to train CroSysLog, referring to ablation studies a1-a3 and b1-b3 in Table \ref{tab:ablation_study}. Both a1 and b1 achieve a high F1-score (a1: 90.84\%; b2: 99.17\%), indicating that the effectiveness of our CroSysLog is not dependent on the current source systems, Liberty and BGL, and it can maintain robust performance with different source system. Besides, a2, a3, b2, b3 only achieve F1-Scores in the range of
33.76\% to 58.71\%, highlighting the importance of using multiple source systems to train CroSysLog with our meta-learning method. This confirms that MAML, as a multi-task meta-learning algorithm, is designed to leverage diverse tasks during training, enabling better adaptation to new contexts \cite{finn2017model}.

Additionally, we assess the contribution of meta-learning by removing MAML algorithm from CroSysLog and using only the base model LSTM under the same experimental settings in each experiment, referring to ablation studies a4 and b4 in Table \ref{tab:ablation_study}. The results show that, without the MAML algorithm, the base model LSTM performs significantly worse,  as evidenced by the low F1-Scores achieved by a4 (17.98\%) and b4 (0.04\%). This confirms the importance of meta-learning in CroSysLog for enhancing the base model's adaptability to target systems for effective anomaly detection.

\begin{table}[!ht]
\caption{Ablation studies in each experiment}
\label{tab:ablation_study}
\centering
\scriptsize
\begin{tabular}{l l rrr}
\toprule
%E1 &&\\
E1 & Training context & \scriptsize Precision & \scriptsize  Recall & F1-Score  
%ab2. CroSysLog & LIBERTY+BGL→SPIRIT & 98.38 & 99.99 & 99.17 \\
%ab3. LSTM & & 
\\ \hline 
a1.CroSysLog & Liberty+Spirit→\newline Thunderbird & 90.38 & 91.73 & 90.84  \\
a2.CroSysLog & Liberty→Thunderbird & 50.03 & 33.57 & 33.76 \\
a3.CroSysLog & BGL→Thunderbird & 63.41 & 59.31 & 58.71 \\
a4.LSTM & Liberty+BGL→\newline Thunderbird & 10.33 & 100.0 & 17.98 \\ \midrule

E2 & Training context & Precision & Recall & F1-Score \\ \hline 
b1.CroSysLog & Thunderbird+BGL→Spirit & 98.38 & 99.99 & 99.17 \\
b2.CroSysLog & Liberty→Spirit & 52.49 & 37.23 & 36.88 \\
b3.CroSysLog & BGL→Spirit & 68.00 & 49.43 & 50.81 \\
b4.LSTM & Liberty+BGL→Spirit & 61.42 & 0.02 & 0.04\\  \hline 
\end{tabular}
\vspace{-8pt}
\end{table}

\section{THREATS TO VALIDITY}
One internal threat may lie in baseline development. Several baseline approaches were originally designed for log-session level anomaly detection. We modified these approaches for log-event level anomaly detection (Section \ref{sec:baselines}) to meet our study context. To mitigate this threat, we carefully studied the relevant methodology, consulted domain experts, and conducted extensive testing. Another internal threat is related to how we prepared meta-testing tasks for each target system: for each meta-testing task, we randomly selected a log split as the support set for task-specific adaptation. Although our evaluation shows good performance, if logs evolve significantly within target systems, selecting a log split as the support set that is closer in time to the log split of the query set, based on the concept of MAML \cite{finn2017model}, could potentially improve results. Also, we utilized two meta-learning tasks of two source systems in each experiment. Initially, we began with one task with one source system, and observed that two source systems with two tasks yielded sufficient performance for the scope of our study. In other contexts, a greater number of meta-learning tasks and source systems might be required to achieve optimal results, according to the nature of  MAML~\cite{finn2017model}.

 The external threat may arise from not yet evaluating CroSysLog on a broader range of software systems, e.g., web service systems, distributed databases.  Due to our limited resources, we have only evaluated CroSysLog on supercomputing systems with open datasets. However, by leveraging MAML, CroSysLog should be highly adaptable and could potentially be adapted to perform anomaly detection in a broader range of new and diverse systems. Expanding the evaluation to include a wider variety of systems would enhance the applicability, generalization, and scalability of  CroSysLog in real-world practices. Another external threat is that we have not yet deployed and tested  CroSysLog in real-world operation and maintenance contexts. Such field trials would provide valuable insights into CroSysLog’s return on investment (ROI), such as its ability to detect anomalies early, enhance operational efficiency, and reduce false positives for organizations that run multiple software systems. 
 
% By leveraging MAML, CroSysLog should be highly adaptable:ould potentially be adapted to perform anomaly detection in a broader range of new and diverse distributed systems. However, the MAML algorithm is designed to be highly adaptable; once trained on multiple tasks with diverse data, it can quickly adjust to new tasks in different contexts with only a few samples \cite{finn2017model}. This suggests that our study may not fully leverage MAML's advantages. By leveraging MAML,  after being trained on log-entry level anomaly detection for multiple software systems with varied log data, our framework could potentially be adapted to perform anomaly detection in a broader range of new and diverse distributed systems. Expanding the evaluation to include a wider variety of systems would enhance the applicability, generalization, and scalability of our framework in real-world AIOps practices. Another external threat is that we have not yet deployed and tested our framework in real-world operation and maintenance scenarios of varied software systems.  Conducting such field trials would allow us to better measure the framework's return on investment in terms of, e.g., early anomaly detection, increased operational efficiency, reduced false positives, and more effective resource allocation.

\section{Conclusion}
%Our empirical study on existing AIOps tools (Section \ref{sec:empirical_study})
%indicates that no single tool is designed to address the identified three practical challenges regarding log-based anomaly detection in real-world operational and maintenance contexts. Thus, 
This paper proposes an AIOps tool, CroSysLog, for log-event-level anomaly detection, specifically responding to practical challenges in real-world operational and maintenance contexts.  CroSysLog can be trained on source systems with sufficient labeled log events, after training, it can efficiently adapt to different target systems with a few labeled log events for effective anomaly detection. It has been evaluated on open datasets of large-scale distributed supercomputing systems demonstrating its performance.
Future work will further address identified threats. We plan to deploy and test CroSysLog on different software systems where logs evolve significantly in real-world operation and maintenance scenarios to observe its ROI.

%Such field trials would provide valuable insights into CroSysLog’s return on investment, such as its ability to detect anomalies early, enhance operational efficiency, reduce false positives. 

%\noindent \textbf{Data Availability} We will publish our replication package upon acceptance. The anonymous link for review: https://figshare.com/s/8fb3739dce4ecaef180b

\section*{Acknowledgment}
This work is funded by the Research Council of Finland (Decision No. 349487) under the academic project MuFAno. The authors acknowledge CSC-IT
Center for Science, Finland, for providing computational resources.

\bibliographystyle{IEEEtran}
\bibliography{own}

\end{document}